\documentclass[10pt,twocolumn,letterpaper]{article}

\usepackage{iccv}
\usepackage{times}
\usepackage{epsfig}
\usepackage{graphicx}
\usepackage{amsmath}
\usepackage{amssymb}
\usepackage{subcaption}
\usepackage[ruled,vlined]{algorithm2e}
\usepackage{makecell}

\usepackage[pagebackref=true,breaklinks=true,letterpaper=true,colorlinks,bookmarks=false]{hyperref}

\iccvfinalcopy % *** Uncomment this line for the final submission

\def\iccvPaperID{7010} % *** Enter the ICCV Paper ID here

\ificcvfinal\fi

\begin{document}

%%%%%%%%% TITLE
\title{Universal and Flexible Optical Aberration Correction Using\\ Deep-Prior Based Deconvolution}

\author{
Xiu Li$^1$\thanks{lixiu15@mails.tsinghua.edu.cn}\ ,
Jinli Suo$^1$,
Weihang Zhang$^1$,
Xin Yuan$^2$,
Qionghai Dai$^1$\\\\
$^1$Tsinghua University,
$^2$Westlake University}

\maketitle
\ificcvfinal\thispagestyle{empty}\fi

%%%%%%%%% ABSTRACT
\begin{abstract}
   High quality imaging usually requires bulky and expensive lenses to compensate geometric and chromatic aberrations. This poses high constraints on the optical hash or low cost applications. Although one can utilize algorithmic reconstruction to remove the artifacts of low-end lenses, the degeneration from optical aberrations is spatially varying and the computation has to trade off efficiency for performance. For example, we need to conduct patch-wise optimization or train a large set of local deep neural networks to achieve high reconstruction performance across the whole image. In this paper, we propose a PSF aware deep network, which takes the aberrant image and PSF map as input and produces the latent high quality version via incorporating deep priors, thus leading to a universal and flexible optical aberration correction method. Specifically, we pre-train a base model from a set of diverse lenses and then adapt it to a given lens by quickly refining the parameters, which largely alleviates the time and memory consumption of model learning. The approach is of high efficiency in both training and testing stages. Extensive results verify the promising applications of our proposed approach for compact low-end cameras. The code is available at \url{https://github.com/leehsiu/UABC}
\end{abstract}

%%%%%%%%% BODY TEXT
\section{Introduction}
\begin{figure}[!t]
\begin{subfigure}{0.28\textwidth}
\centering
\includegraphics[width=\linewidth]{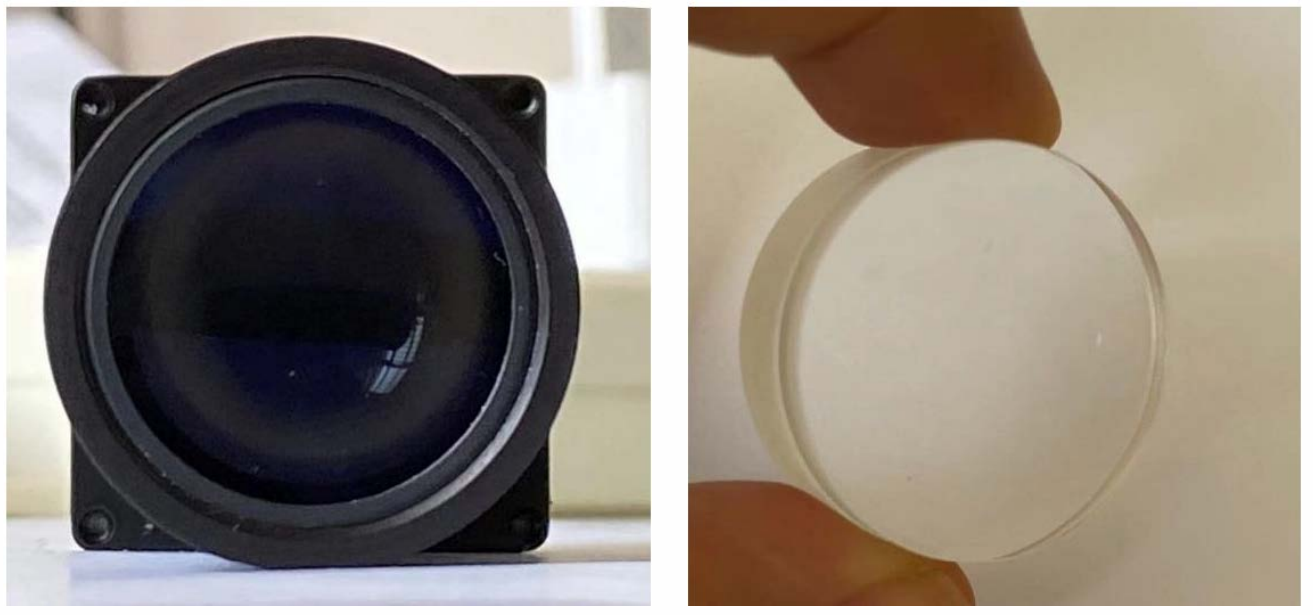}
\caption{}
\end{subfigure}~
\begin{subfigure}{0.187\textwidth}
\centering
\includegraphics[width=0.9\linewidth]{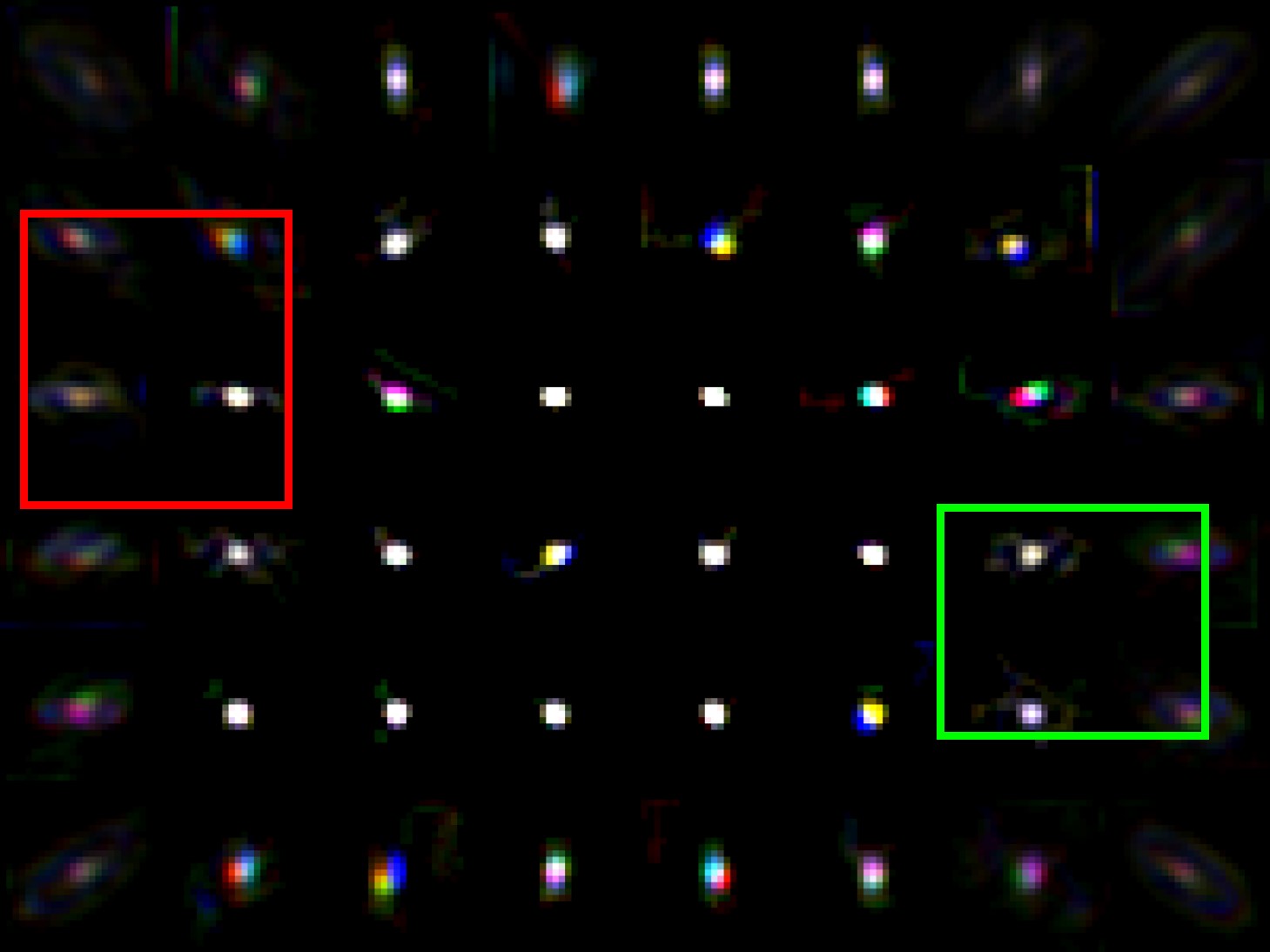}
\caption{}
\end{subfigure}

\begin{subfigure}{\linewidth}
\centering
\includegraphics[width=0.98\linewidth]{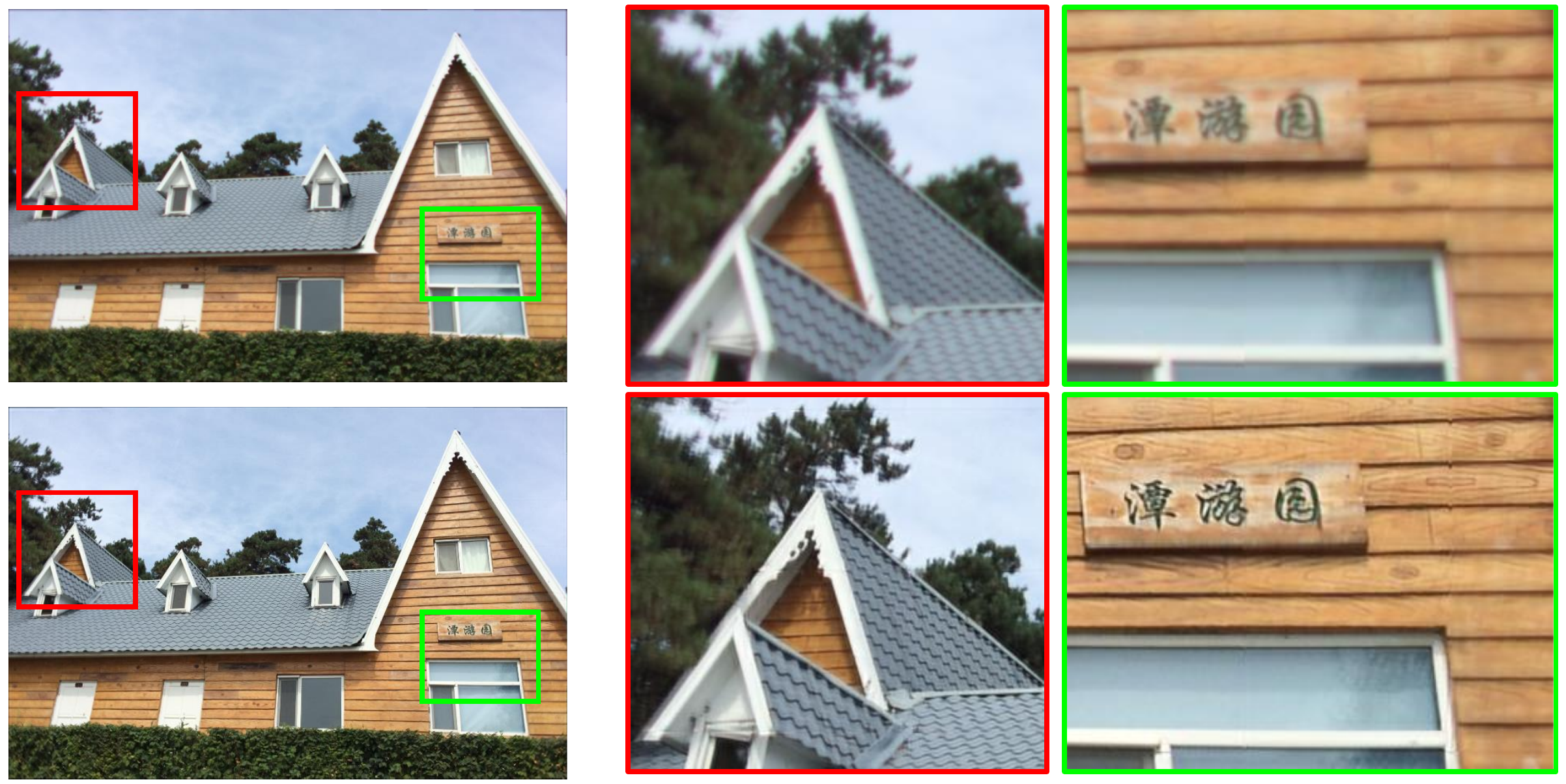}
\caption{}
\end{subfigure}
\caption{One example of computationally reconstructing high quality image with a simple lens. (a) A camera with a simple double glued lens (Thorlabs, AC254-075-A-ML). (b) The calibrated PSF of the camera in (a). (c) The input degenerated image  (upper row) and our reconstruction result (bottom row).}
\label{fig:teasor}
\end{figure}

Optical aberration is one of the most common degeneration in real lens-based imaging systems. Due to the deviation from ideal thin-lens model, the simple/single lens elements suffer from chromatic, spherical aberration and coma aberrations, and degenerate the imaging quality significantly. To cancel out these artifacts, modern camera lenses are usually made of a complex combination of several (even dozens of) lens elements with carefully designed parameters (aka. lens prescription). In a nutshell, existing techniques achieve high imaging quality via such complex design, at the expense of high cost, bulky weight and inevitable lens flare. With the rapid development of large pixel count digital sensors (\eg 100 mega-pixel scale), effective compensation of lens aberration is highly desired.

To achieve \textit{light-weighted} and \textit{low-cost} high-quality imaging, computational optical aberration correction has been exploited during the past decades. Different from advanced optical design, computational methods employ a simple lens for imaging and remove the aberration afterwards with algorithms. Mathematically, the lens aberration can be formulated as convolution with spatially varying kernels, and the compensation is conducted by deconvolution with the assistance of image priors. These computational methods can be either \textit{non-blind} or \textit{blind} depending on whether the aberration model, \ie, Point Spread Function (PSF) is calibrated beforehand. The former measure the PSF using specially designed systems~\cite{ImageEnhance,HeideImaging,KeeDeblur} or estimating it from degenerated images using some developed algorithms explicitly~\cite{SchulerNonStation,TangSingleCalibration,patchdeblur_iccp2013}, deconvolution is conducted afterwards. The latter usually jointly estimates the PSF and the latent sharp image in an iterative manner~\cite{taoNonuniform}. Considering the fact that the PSF is fixed for a given lens, and the estimation of PSF has already been extensively studied, we focus on the non-blind case in this paper. 

As illustrated in Fig.~\ref{fig:teasor}(b), the PSF by lens aberration is spatially non-uniform and varying across color channels as well. Previously, researchers have proposed a series of algorithms for non-uniform deconvolution, includes pixel-wise deconvolution~\cite{nonunidensity,WhyteNonuniform,FergusShaken} and patch-wise deconvolution~\cite{HeideImaging,SchulerNonStation,YueRadial,HirschFast}. The later one assumes that the PSF varies smoothly in the spatial dimension thus can be approximated locally uniform. The deconvolution in this situation is usually faster than the pixel-wise methods. This assumption also holds for the PSF caused by the aberration of a low-end lens. 

Recently, deep neural network has also been exploited for image deconvolution and has shown encouraging results~\cite{fastDeepDeconv,deepWiener}. However, handling non-uniformity is still challenging for end-to-end deep learning. Using a single network weight setting for different sub-regions with different PSF will either requires a large model capacity or leads to an average performance. While training a series of models specifically for each sub-region is quite time consuming. In this paper, we propose to combine deep network and model-based deconvolution into an iterative framework. Prior-guided deconvolution is conducted locally and explicitly models the physical degeneration process, which ensures the high fidelity and easy adaption to different lenses. A single global deep projection network is applied to the whole image to suppress ringing artifacts and eliminate blocking artifacts. Meanwhile, we propose to pretrain a base model from a large set of diverse lenses first and then adapt it to a specific lens quickly, greatly simplify the training process for different lenses.

The main contributions of this paper are as follows.
\begin{itemize}
    \item We propose a PSF-aware deep network to address the spatially varying degeneration caused by lens aberration,  by {\em decoupling} the physical imaging model and deep network prior.
    \item The model training is built on a pre-trained base model plus a fast {\em adaption} procedure to different lenses. Thus, it is of high efficiency and high feasibility.
    \item The approach achieves state-of-the-art performance but with much higher running efficiency.
\end{itemize}
We conduct synthetic and qualitative experiments to demonstrate the training efficiency and performance of the proposed method. We also get promising results on real data captured by low end lenses,  please refer to Fig.~\ref{fig:teasor}, and more results in Fig.~\ref{fig:RealData}.

\section{Related work}
\subsection{Computational aberration correction}
Computational aberration correction can be roughly divided into non-blind and blind methods. For non-blind method, an explicit calibration stage is required to estimate the lens aberration model. Shih et al.\cite{ImageEnhance} literately measures the point-spread function by imaging a pinhole grid pattern in a dark room. Heide et al.\cite{HeideImaging} calibrate the PSF by using a framed random pattern. Specially designed calibration chart~\cite{JoshiSharp,KeeDeblur,SchulerNonStation} is also widely used due to its lower cost. Once the blur kernel is known, the latent clear image can be solved by image deconvolution~\cite{HeideImaging,SchulerNonStation}. Blind methods skips this calibration stage and directly estimate the blur kernel from the images based on the natural image statistics and knowledge of the aberration model. Rahbar and Faez~\cite{ZenikeCalibration} used a Zernike model to describe the lens aberration and estimated the Zernike polynomial coefficients statistically. Schuler et al.~\cite{SchulerBlind} proposed an Efficient Filter Flow (EFF) basis method to describe the non-uniform blurry kernels by exploiting the reflection symmetry, rotational symmetry and radial distributions of the PSF. This prior knowledge is also used by Yue et al.~\cite{YueRadial} while they also propose a radial-splitting technique (representing the image in polar coordinates) and a sharp-to-blur estimation strategy, which is further extensively studied by\cite{sun2017revisiting}.
\subsection{Image deconvolution}
Mathematically, optical aberration correction is a special case of non-uniform image deconvolution. Usually, the modelling of the non-uniformity can be done fully paramatrically, which results in a pixel-wise deconvolution~\cite{nonunidensity,WhyteNonuniform,FergusShaken,compactHI}. To accelerate the calculation, the PSF can also be assumed to vary smoothly locally, thus the non-uniform deconvolution can approximated with patch-wise uniform deconvolution.~\cite{HeideImaging,SchulerNonStation,YueRadial,HirschFast}. As image deconvolution is usually ill-posed, regularization terms (or priors) are essential to produce reasonable result. General prior includes TV-norm minimization~\cite{TVNorm}, hyper-Laplacian~\cite{LapDeconv,LapDeconvBill} and self-similarity~\cite{RecurNorm}. In the recent years, deep learning has been widely exploited for deconvolution problems for its ability in modelling complex data distribution. Although it is possible to directly learn to map the blurred image to its corresponding clear version~\cite{Zhang_2019_CVPR_deblur_blind,deblurGAN}, is has been shown that non-blind deep deconvolution performs better than direct end-to-end mapping~\cite{Zhang_2017_CVPR_deconv,KruseDeepDeconv,SchulerDeepDeconv,fastDeepDeconv,deepWiener}. These works utilize the PSF by convert the deep deconvolution problem into deconvolution and network based refinement. This way the network is capable to handle various degeneration kernels.

\subsection{Plug-and-Play algorithms}
Another trending way to utilize PSF-awareness is plug-and-play(PnP)~\cite{pmlr-v97-ryu19a,PnPLearning}. A straightforward way is replacing the general or handcraft prior with a learned deep generative model~\cite{DGP} solve the optimization problem using gradient decent. To avoid the heavy computational load of tracking the network gradient, the deep network can also serves as a proximal operator, and the image restoration problem is solved by iteratively running model-based optimization and deep network projection. This modeling has shown promising results in linear inverse problems\cite{OneNet}, for image super-resolution~\cite{Zhang_2020_CVPR}, snapshot compressive imaging~\cite{Yuan20PnPSCI} and high-spectral imaging~\cite{compactHI}. Our work also follows this formulation.

\section{Method}
In this section, we first mathematically formulate the lens aberration model and then propose our deep-prior based approach for optical aberration correction.

\begin{figure}[t!]
    \centering
    \includegraphics[width=\linewidth]{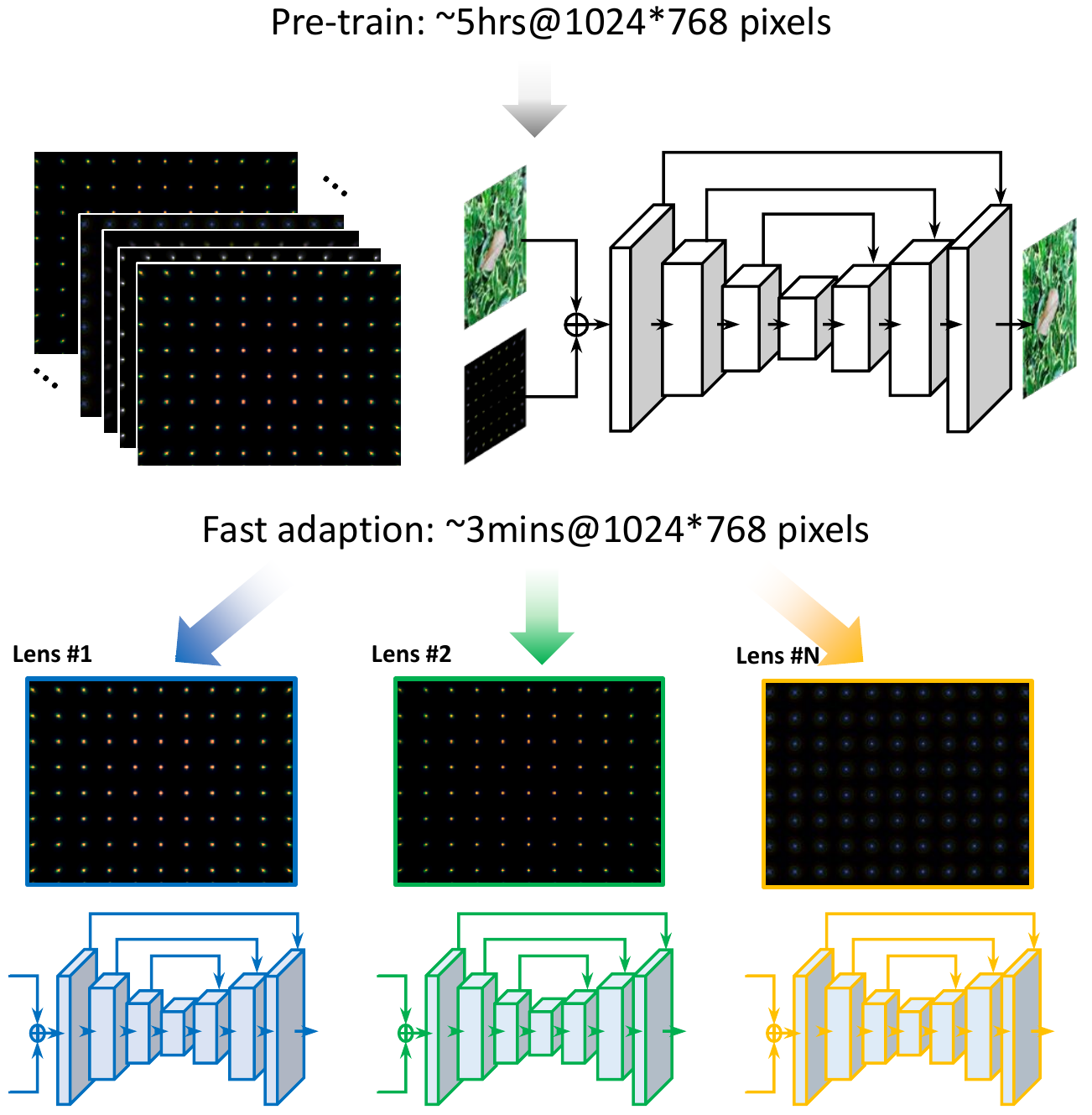}
    \caption{The basic idea of building fast and high quality lens aberration correction model, via pre-training a base model and fast adapting to different lenses.}
    \label{fig:Pnet}
\end{figure}

\subsection{Problem formulation}
As mentioned before, lens aberration can be interpreted as non-uniform channel-wise blur as the PSF varies both spatially and spectrally (\ie, RGB channel). Thanks to the fact the neighbouring PSFs are highly correlated,  we can approximate the non-uniform PSF as patch-wise uniform. 
For each small image patch $p$ at channel $c$, the recorded degraded image patch $\mathbf{y}_{c,p}$ and the latent clear patch $\mathbf{x}_{c,p}$ is related by
\begin{equation}
    \mathbf{y}_{c,p} = \mathbf{k}_{c,p}\otimes \mathbf{x}_{c,p} + \mathbf{n}_{c,p}, 
    \label{Eq:patch_conv}
\end{equation}
where $\mathbf{k}_{c,p}$ is the corresponding PSF (blur kernel), $\mathbf{n}_{c,p}$ is the noise and $\otimes$ denotes the 2D convolution. For simplicity, we will drop this $c$ index if we consider $\otimes$ is carried channel-wisely. 

To solving Eq.~\eqref{Eq:patch_conv}, a prior term is necessary to constrain the space of $\mathbf{x}$. This leads to
\begin{equation}
\mathbf{x}=\arg\min_\mathbf{x} \sum_{p} \lVert \mathbf{k}_p\otimes \mathbf{x}_p-\mathbf{y}_p  \rVert^2_2 + \lambda \Phi({\mathbf{x}}),
\label{Eq:min_x}
\end{equation}
where the first term enforces the data term residual and the second one $\Phi(\mathbf{x})$ denotes the prior(statistical distribution) of $\mathbf{x}$.

\begin{figure*}[htbp!]
    \centering
    \includegraphics[width=\linewidth]{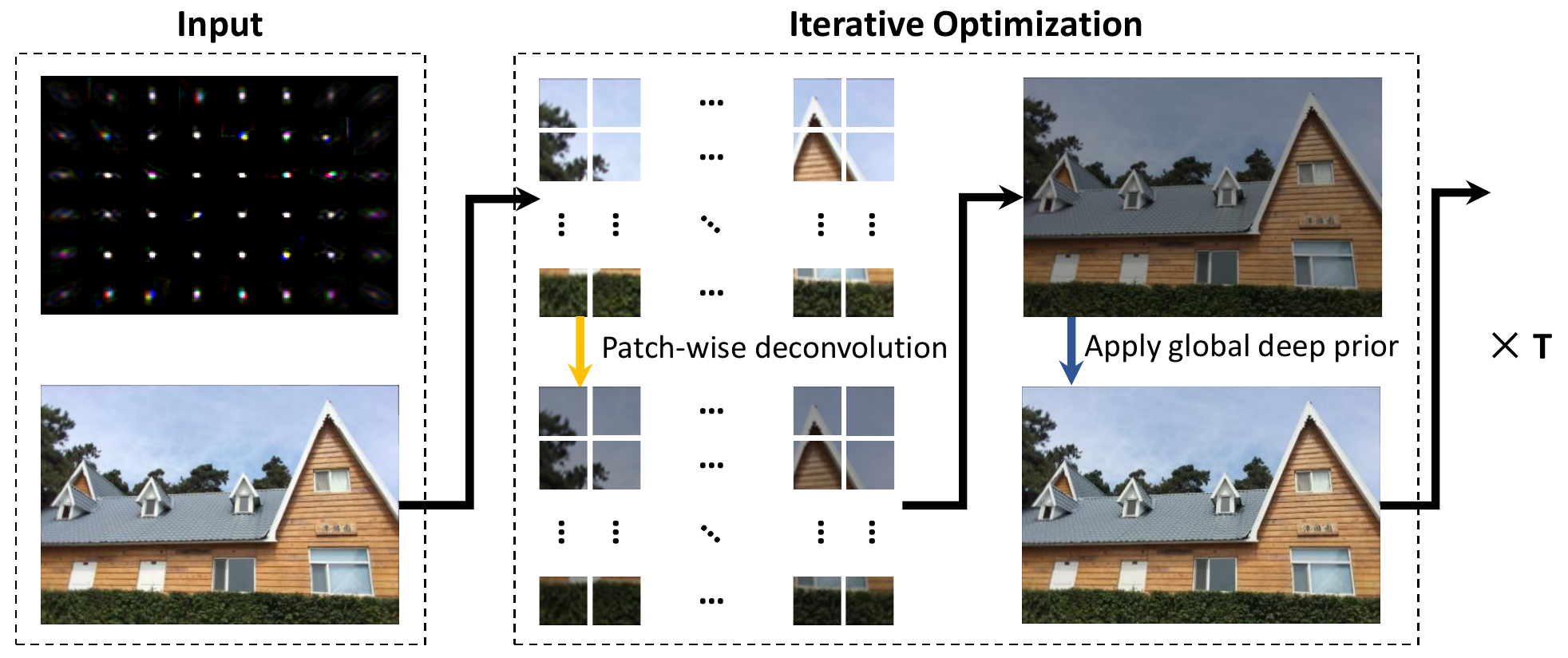}
   % \vspace{-2mm}
    \caption{The flow chart of the proposed approach. The divided two parts---deconvolution and prior imposition, are illustrated with different colors.}
    %\vspace{-3mm}
    \label{fig:pipeline}
\end{figure*}

\subsection{Data/Prior splitting}
We follow ~\cite{compactHI,eboli2020endtoend,Zhang_2020_CVPR} to use Half-Quadratic Splitting (HQS) to solve Eq.~\eqref{Eq:min_x}. By introducing an auxiliary variable $\mathbf{z}$, Eq.~\eqref{Eq:min_x} can be re-modeled as 
\begin{equation}
\arg\min_{\mathbf{x},\mathbf{z}}  \sum_{p}\lVert \mathbf{k}_p\otimes\mathbf{z}_p-\mathbf{y}_p  \rVert^2_2 + \lambda\Phi({\mathbf{x}}) + \mu \lVert \mathbf{z} - \mathbf{x}\rVert^2_2,
\end{equation}
which can be solved by performing iterations of the following sub-problems.
\begin{align}
\mathbf{z}^{t} &\leftarrow \min_\mathbf{z} \sum_p\lVert \mathbf{k}_p\otimes\mathbf{z}_p - \mathbf{y}_p\rVert^2_2 + \mu \lVert\mathbf{z} - \mathbf{x}^{t-1}\rVert^2_2, \label{Eq:z_k}\\
\mathbf{x}^{t} &\leftarrow \min_\mathbf{x} \mu \lVert \mathbf{z}^{t} - \mathbf{x}\rVert^2_2 + \lambda \Phi(\mathbf{x}). \label{Eq:x_k}
\end{align}

Eq.~\eqref{Eq:z_k} can be approximately solved in a closed form under circular boundary conditions as
\begin{equation}
    \mathbf{z}^{t}_p = \mathcal{F}^{-1}\left(\frac{\overline{\mathcal{F}(\mathbf{k})}\mathcal{F}(\mathbf{y}_p) + \mu^{k}_p \mathcal{F}(\mathbf{x}_p^{t-1})} {\overline{\mathcal{F}(\mathbf{k})}\mathcal{F}(\mathbf{k})+\mu_p^{k}}\right),
    \label{Eq:z_kp}
\end{equation}
where $\mathcal{F}(\cdot)$ denotes the Fast Fourier Transform (FFT), $\mathcal{F}^{-1}(\cdot)$ denotes the inverse FFT, and $\overline{\mathcal{F}(\cdot)}$ denotes the complex conjugate of FFT. Notice that although we use a single $\mu$ notation in Eq.~\eqref{Eq:z_kp}, we use different $\mu_p^t$ for different patches, channels and iterations when implementing the patch-wise deconvolution.

For Eq.~\eqref{Eq:x_k}, if we absorb $\mu_{p}^t$ into $\lambda$, it is a problem with $\mathbf{z}_p^{t},\lambda_p^{t}$ as input and $\mathbf{x}^{t}$ as output. We treat $\Phi(\cdot)$ as a proximal operator and directly learn a deep projection network  $\mathcal{P}$  similar to ~\cite{OneNet,Zhang_2020_CVPR,compactHI}.
\begin{equation}
    \mathbf{x}^{t} = \mathcal{P}(\mathbf{z}^{t}_p,\lambda_p^{t}).
\end{equation}

Note that we have omitted the chop and assemble operations between $\mathbf{x},\mathbf{z}$ and $\{\mathbf{x}_p\},\{\mathbf{z}_p\}$

\subsection{Deep projector $\mathcal{P}$}
The deep projector $\mathcal{P}$ is built mainly following the U-net~\cite{UNet}, which is widely used for image-to-image translation problems. Here we replace the original plain convolution in original U-net with residual blocks for better convergence. Notice that as $\mathcal{P}$ is fully convolutional, it can be applied to images with an arbitrary size. Thus unlike the deconvolution stage which is done patch-wisely, we ensemble the patches together and update the whole image in this stage. The input of $\mathcal{P}$ is 6 channels(by assembling and concatenating $\mathbf{z}^t_p$ and $\lambda^t_p$) and output is 3 channels $\mathbf{x}$ with the same spatial dimension. The network structure is shown in Fig.~\ref{fig:Pnet}. Ringing artifacts and blocking effects caused by the patch-wise deconvolution are effectively removed because information of neighboring patches will be used. The flow of the full forwarding process is illustrated in Fig.~\ref{fig:pipeline}.

\begin{algorithm}[t]
\SetAlgoLined
 \caption{Framework of our proposed deconvolution method.}
\KwIn{ $\mathbf{y}$, $\mathbf{k}$, max stage number $T$, $\{\mu_p^t\}$,$\{\lambda_p^t\}$.}
 \BlankLine
 {\bf Pre-processing}\;
  1. Chop the input blur image into non-overlapping patches according to $\mathbf{k}$ and padding each patch with pixels from neighboring patches;\\
  2. Pre-calculate $\{\mathcal{F}(\mathbf{k}_p)\},\{\mathcal{F}(\mathbf{y}_p)\}$. \\
  3. Initialize $\mathbf{x}^0$ as zero, thus the first stage update of $\mathbf{z}_{p}$ is the same as the Wiener filter.\;
 
 \For{$t=1,...,T$}
 {1. Chop $\mathbf{x}^{t-1}$ into patches using same strategy in the pre-processing stage, and apply Eq.~\eqref{Eq:z_kp} to get $\{\mathbf{z}_p^t\}$.\\ 
 2. Shave $\{\mathbf{z}_p^t\}$ and assemble them together to get $\mathbf{z}^t$. Then pass $\mathbf{z}^t$ and $\mathbf{\lambda}_t$ through the $\mathcal{P}$ to get $\mathbf{x}^t$. 
 }
 \KwOut{$\mathbf{x}^T$}
 \label{alg:algorithm1}
\end{algorithm}

\begin{figure*}[t!]
    \centering
    \includegraphics[width=.98\linewidth]{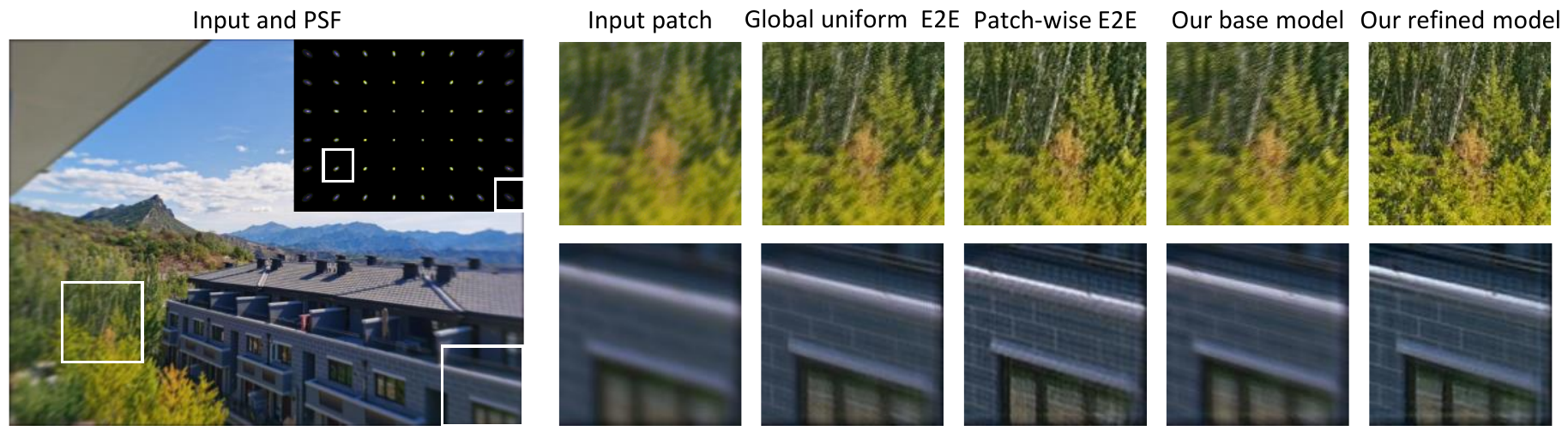}
    \vspace{-1mm}
    \caption{Comparison of visual results among different deep learning models, in correspondence to those in Table.~\ref{tab:training_eff}.}
    \vspace{-2mm}
    \label{fig:DiffSet}
\end{figure*}

\subsection{Pretrain and Adaption}

\noindent\textbf{End-to-end Training} 
We use DIV2K and Flicker2K dataset~\cite{DIV2K}, with 3450 high-definition high-resolution ($2K$) images in total, to train our model. We synthesize  the training triplets (blurred image $\mathbf{y}$, aberration PSF $\mathbf{k}$ and clear image $\mathbf{x}$) as follows. The ground truth clear image $\mathbf{x}$ is drawn from the training dataset. The PSF $\mathbf{k}$ is generated by two ways: (i) We use Zemax software to simulate the PSF for various simple-lens designs. The kernel size of the PSF is 25\texttimes 25 and the spatial resolution is set to 16\texttimes 16 (in total, we have 256 different PSFs for each lens prescription). (ii) Anisotropic Gaussian kernels with different deviations are also used to increase the PSF diversity. The degenerated image $\mathbf{y}$ is generated via Zemax software by ray tracing (for PSF wth known prescriptions) or patch-wise convolution (for Gaussian kernels).

\begin{figure}[htbp!]
    \centering
    \includegraphics[width=\linewidth]{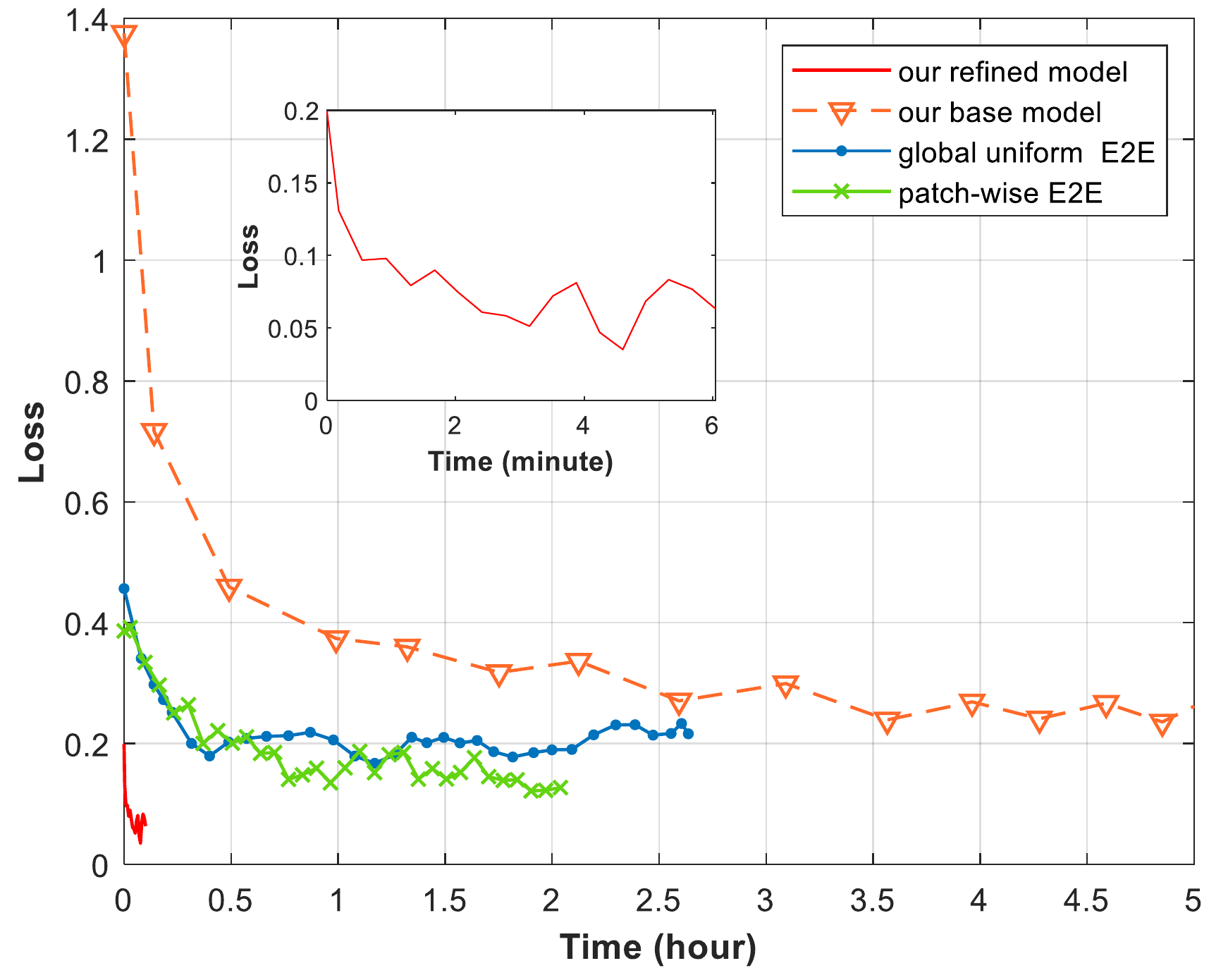}
    \caption{Training curves of four different learning strategies, in correspondence to those in Table~\ref{tab:training_eff}. Here the curve for the patch-wise E2E is the average training time for a single patch.}
    \label{fig:trainingCurve} 
\end{figure}

Directly training on the full image with spatially-varying PSF is infeasible due to the GPU memory limits. Instead, for each training sample we randomly crop the image with 256\texttimes 256 patch size, and also randomly crop a 2\texttimes 2 patch from the whole 16\texttimes 16 PSF map of a randomly drawn lens prescriptions. The blurred image thus has four regions with different PSFs. The network is trained by minimizing the $\ell_1$ reconstruction error. The batch-size is set to 3, resulting in a 10.6G memory cost on a single GPU during training. The $\mu,\lambda$ are set as free trainable parameters optimized together with the $\mathcal{P}$ network parameters using Adam optimizer~\cite{kingma2015adam} with the $0.001$ learning rate. Altough we mentioned above that $\mu,\lambda$ are set different for each sub-region, in the pretrain stage, they are set as spatially uniform, only varies across different stages. It takes around 5 hours to train the model for 7K iterations using PyTorch implementation on a single 12G memory Nvidia RTX 2080 Ti GPU. One exemplar training curve is shown in Fig.~\ref{fig:trainingCurve}. 

Notice that for each training iteration, the network will see the assembled patches with different PSFs, thus the network needs to learn to remove the deconvolution artifacts but also the blocky artifacts. This implementation is crucial to the success of our methods.

\noindent\textbf{Kernel-specific refine} When a specific PSF is presented, we can refine the above pre-trained base model for performance boosting. We follow the same training protocol in the end-to-end training stage. Differently, instead of optimizing a spatially uniform $\mu,\lambda$ variable, we built a $\mu,\lambda$ map with each one corresponding to a specific local PSF. Experiments show that after lens specific adaption, the performance will raise significantly even with a very small number of training iterations, \eg, 3 minutes for around 60 iterations.
For the lenses with more severe aberrations in the marginal regions, we will use smaller patches in the outer regions than in the central ones. 

\section{Results}

\begin{table}
    \centering
        \caption{Comparison of efficiency and performance of different strategies. For each comparison, best results are highlighted and second best one emphasized in bold font.}
        \vspace{-1mm}
    \resizebox{.48\textwidth}{!}
    {
    \begin{tabular}{c|c|c|c}
        \hline
        Approach & Training time & PSNR (dB) & SSIM\\
        \hline
        {Global uniform E2E} & $\sim$ 2 hrs &{26.47} & 0.9106\\
        \hline
        {Patch-wise E2E} & $\sim$ 90 hrs & 27.01 & 0.9110\\
        \hline
        Our base model & $\sim$ 5 hrs & 26.27 & {0.9127}\\
        \hline
        Ours refined model& $\sim$ \textbf{3 mins} & \textbf{27.36} & \textbf{0.9382}\\
        \hline
    \end{tabular}
    }
    \label{tab:training_eff}
\end{table}

\begin{figure}[htbp!]
    \centering
    \includegraphics[width=\linewidth]{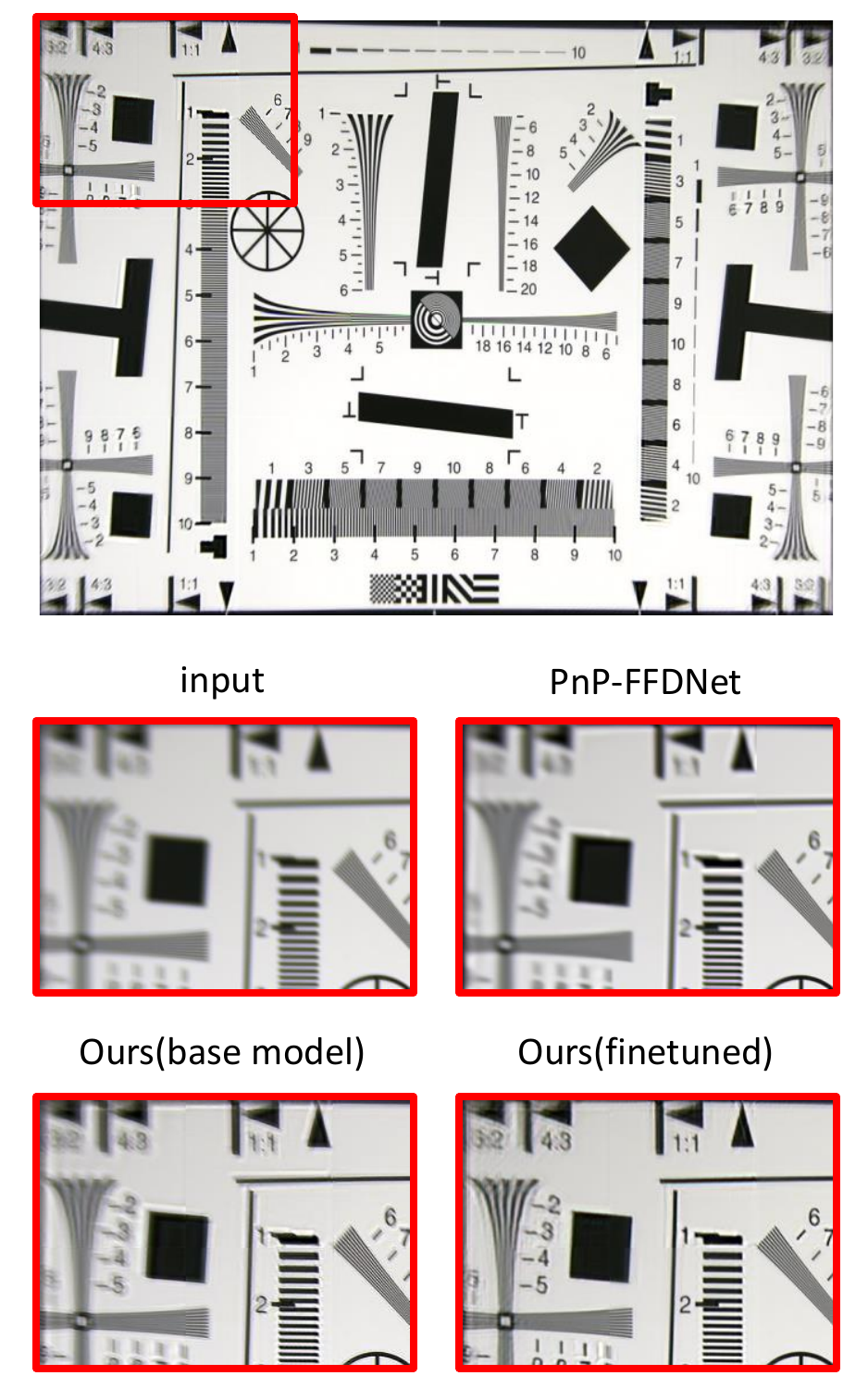}
    \caption{Comparison among different PnP schemes, with general denoiser (FFDNet), universal lens aberration compensation network (our base model) and lens specific compensation network (our refined model), incorporated in the same iterative framework.}
    \label{fig:CmpPNP}
\end{figure}

\begin{figure}[htbp!]
    \centering
    \includegraphics[width=\linewidth]{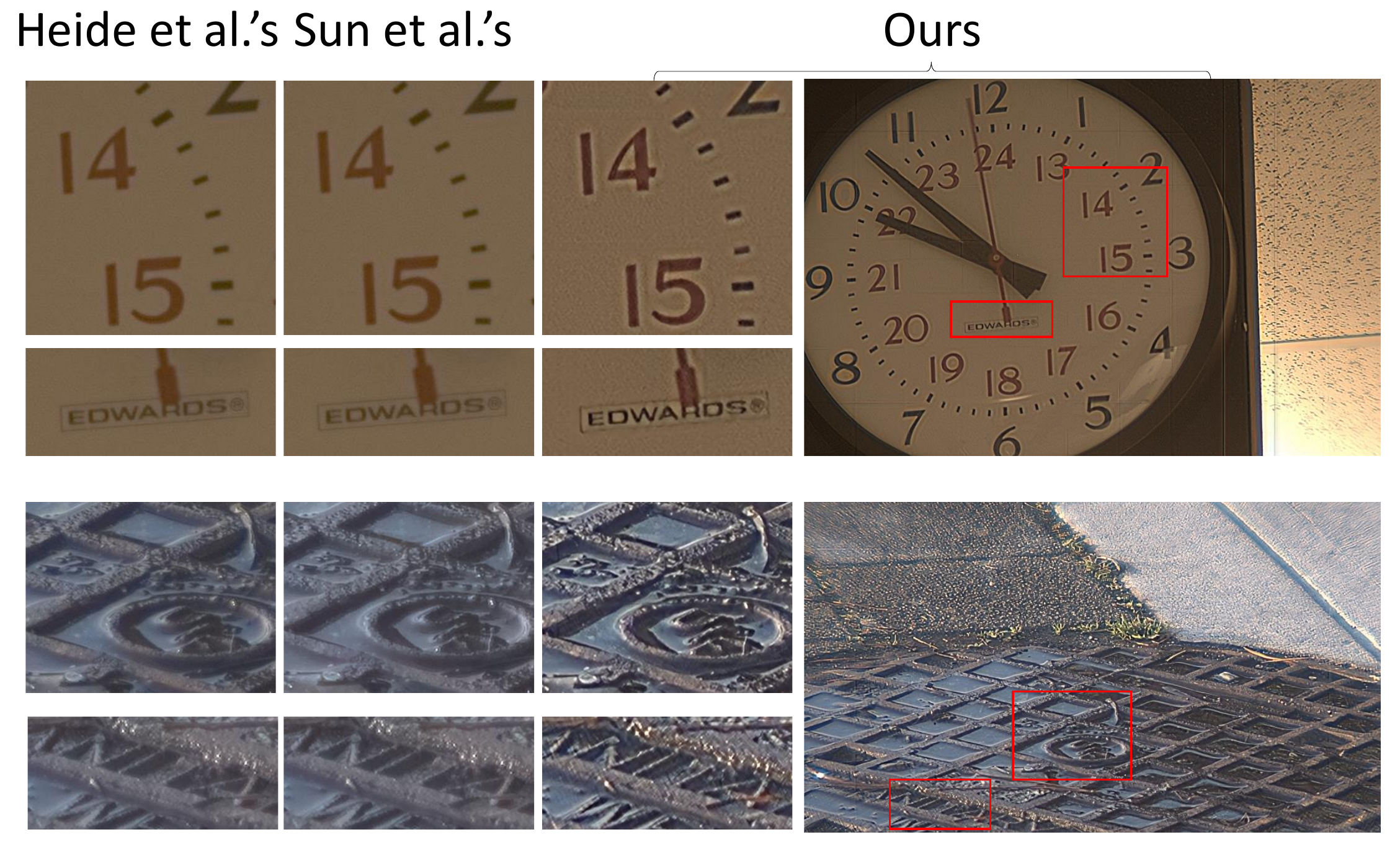}
    \caption{Comparison with one recently published method proposed by Sun et al.~\cite{sun2017revisiting}.}
    \label{fig:CmpSTAR_iccv}
\end{figure}

\begin{figure*}[t]
    \centering
    \includegraphics[width=\linewidth]{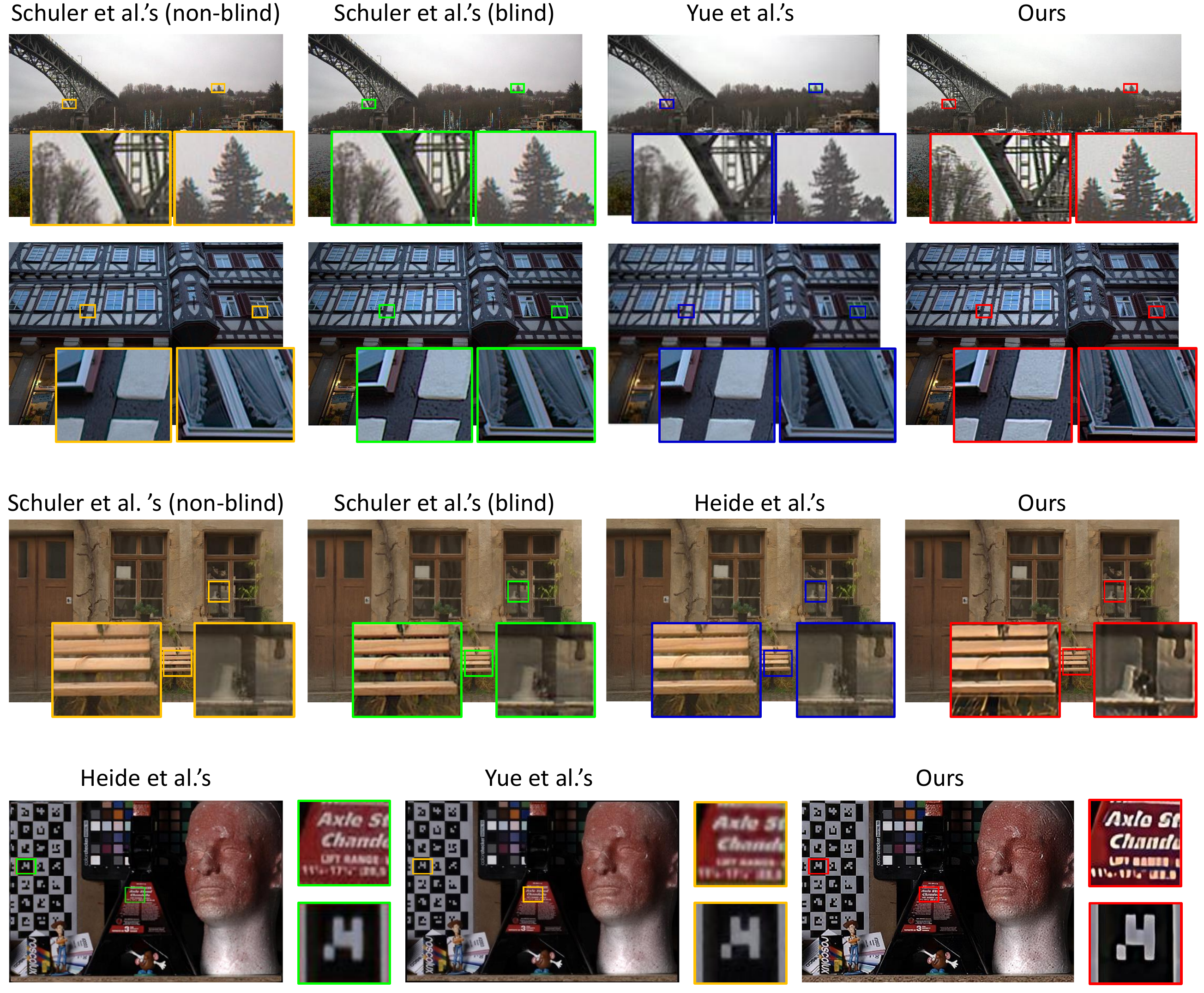}
    \vspace{-3mm}
    \caption{Comparison with state-of-the-art methods, including Schuler et al.'s~\cite{SchulerNonStation}\cite{SchulerBlind}, Yue et al.'s~\cite{YueRadial}, and Heide et al.'s~\cite{HeideImaging}.}
    \vspace{-2mm}
    \label{fig:CmpSTAR}
\end{figure*}

\subsection{Quantitative evaluation}\label{sec:quantExp}
To demonstrate the efficiency and performance of our proposed scheme, we compare our model with three baseline lens aberration compensation methods based on deep learning: ($i$) globally uniform end-to-end (E2E) model by assuming the PSF being uniform and train an single E2E model applying on the full image, ($ii$) non-uniform patch-wise E2E model, in which we decompose the image into PSF-approximately-uniform  patches and train one model for each image patch, and ($iii$) our pre-trained base model, \ie, train a non-uniform model working for a set of diverse lenses. For efficiency evaluation, we compare the time for model training and inference for all the methods. We use DIV2K~\cite{DIV2K} validation set as the testing data and evaluate the restoration results in terms of peak-signal-to-noise-ratio (PSNR) and SSIM~\cite{Wang04imagequality}.

For our base model, we use synthetic aberrant-sharp image pairs and PSFs of 20 different lenses with known Zemax prescriptions. For the other three settings, we use three simple lenses from Thorlabs Corp.: AC254-075-A-ML, LA4924-A, and LA5714. During the experiment, the synthetic image is set to 1024 $\times$ 768 pixels, and the patch size is set to be 128 $\times$ 128 pixels, under which the PSF can be roughly assumed uniform within a patch. The experiments are run on a computer with GTX TITAN 12G Memory GPU, Intel i9 CPU and 16G RAM. 

The results are shown in Table~\ref{tab:training_eff} and one visual comparison is shown in Fig.~\ref{fig:DiffSet}, from which we can see that assuming the PSF to be globally uniform will cause low reconstruction quality, which again verifies that non-uniform compensation is crucial for lens aberration. Among the experiment settings applied to non-uniform deconvolution, our final results achieve comparable performance with the patch-wise E2E method, and superior performance to the base model before lens-specific refining. 

Regarding time to train a model for a given lens, the patch-wise E2E model is exhausting since the model learning for the patches with severe degeneration is quite time consuming. and we usually need to learn tens of models for a mega-pixel image. For the globally uniform model, the required time is similar to that for learning a local E2E model. For the base model working generally for most lenses, around 5 hours are required for training, while for our lens specific adaption, one can easily build a high-quality aberration removal model on the top of the pre-trained base model within minutes, see the training curve in Fig.~\ref{fig:trainingCurve}. In the inference stage, the deep learning based methods take similar time, around \textbf{2.4 seconds} for a \textbf{1024$\times$768} input image under our settings.

Overall, we show that non-uniform PSF awareness is a essential factor in solving the optical aberration correction problem using deep learning. Our universal pretrain plus lens-specific finetune strategy is also of both high efficiency and  performance, making it a practical method.

\begin{figure*}[htbp!]
    \centering
    \includegraphics[width=\linewidth]{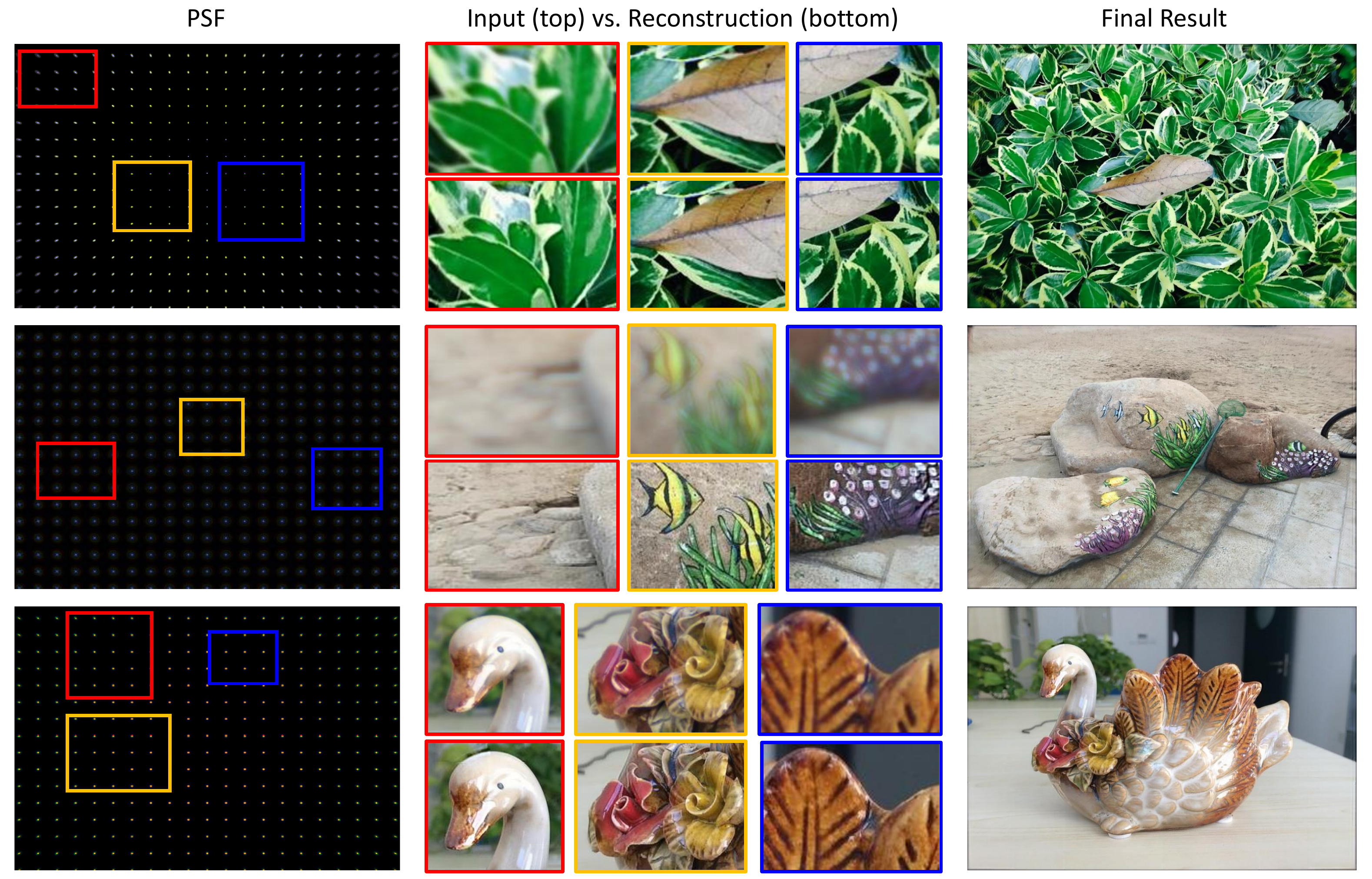}
    \vspace{-1mm}
    \caption{Results on real data captured with 3 different simple lenses. Left: estimated PSF. Middle: the comparison between the input and aberration-compensated version of some regions. Right: the holistic image of the final reconstruction. Please see more result in the supplemental material.}
    \label{fig:RealData}
\end{figure*}

\subsection{Comparison to other plug-and-play schemes}
Plug-and-play is an effective option for image restoration under non-uniform degeneration. Some recent researches treat the prior imposition as a denoising operation and use a generic deep image denoiser for different tasks. Therefore, one can also pre-train a denoising network and incorporate it into our iteration framework of Eq.~\ref{Eq:z_k} and Eq.~\ref{Eq:x_k} in a plug-and-play fashion. Besides, if we do not conduct lens-specific refinement, we can directly use the base model to any lens aberration model, as the plug-and-play method does.

To demonstrate the superiority of building aberration specific model and conducting lens-specific optimization, we conduct a comparison with above two plug-and-play schemes. For the generic denoiser we use the FFDNet~\cite{FFDNet}. We test different noise levels and find the one with best performance for the noise parameter setting in plug-n-play framework. The results are shown in Fig.~\ref{fig:CmpPNP}, where we used the standard resolution chart for test and use the PSF of the same lens as in Fig.~\ref{fig:teasor}.  We found that PnP-FFDNet produces more blurry results than our base model and the lens-specific model. This is reasonable because the deconvolution artifacts might be largely different from the Gaussian noise. The refined network performs better than the base model without lens-specific adaption. This validates the necessity of adaption.

\subsection{Comparison to state-of-the-arts}
We compare our method with other state-of-the-art aberration correction methods including Schuler et al.'s~\cite{SchulerBlind}~\cite{SchulerNonStation}, Yue et al.'s~\cite{YueRadial}, Heide et al.'s~\cite{HeideImaging}, and Sun et al.'s~\cite{sun2017revisiting} on real captured data. The results are presented in Figs.~\ref{fig:CmpSTAR_iccv} and \ref{fig:CmpSTAR}. As there is no ground-truth data, we only compare them qualitatively. It can be observed that: (i) our deep learning based methods have a superior or comparable performance over those generic prior optimization-based methods, which indicates that the deep model can represent the statistics of nature images quite better; (ii) blind compensation can achieve comparable performance with the non-blind ones, if PSF can be reliably estimated from the input degenerated image, and (iii) our approach performs better than (at least comparable to) state-of-the-arts and is quite promising, considering its high flexibility and efficiency.

\subsection{Results on real data}
To test the performance of our approach on real data, we mount the simple lenses on a digital camera (HIKROBOT MV-CB013-20UC-C) to capture some photos, and conduct computational lens aberration. Here we demonstrate three cases with the same simple lenses as in Sec.~\ref{sec:quantExp}. We use the algorithm by Sun et al.~\cite{patchdeblur_iccp2013} to estimate the PSFs from images of several printed binary patterns and taking average to raise the performance, as shown in the leftmost column in Fig.~\ref{fig:RealData}. We then finetune our pretrained base model to each lens, and then feed the calibrated PSF together with the input degenerated image into our framework for final reconstruction, in the rightmost column in Fig.~\ref{fig:RealData}. We can see that the proposed approach can remove the blur effectively for different lenses, supporting our proposed method to be `universal' and `flexible'.

\section{Summary and discussions}
This paper reports a computational reconstruction approach using the end-to-end deep neural network to achieve high-quality imaging with a simple lens. Technically, the model is of high efficiency in both training and inference, and of high flexibility for adapting to different lenses. 

Regarding the future work, we plan to build a training database from a customized lens set that covering tens of Zernike polynomial terms and learn a universal base model that can be adapted faster. Extending the model to explicitly incorporate cross-channel correlation and thus improving the final performance is another ongoing work.

\section*{Acknowledgements}
This work is jointly funded by Ministry of Science and Technology of China(Grant No.2020AA0108202), National Natural Science Foundation of China(Grant No.61931012, 61627804 and 62088102), and Beijing Municipal Natural Science Foundation (Grant No.Z200021). 

{\small
\bibliographystyle{ieee}
\bibliography{ref}
}

\end{document}